\documentclass[10pt]{article}
\usepackage{fullpage}
\usepackage{amsmath}
\usepackage{amssymb}
\usepackage{amsfonts}
\usepackage[dvips]{epsfig}
\usepackage{color}
\usepackage{cite}

\def\##1{{\bf #1}}
\def\=#1{\underline{\underline{#1}}}

\def\+#1{\underline{\bf #1}}
\def\*#1{\underline{\underline{\bf #1}}}

\def\r#1{(\ref{#1})}
\def\l#1{\label{#1}}
\def\c#1{\cite{#1}}

\def\le{\left(}
\def\ri{\right)}
\def\les{\left[}
\def\ris{\right]}

\def\.{\mbox{ \tiny{$^\bullet$} }}

\def\eps{\varepsilon}

\def\epso{\eps_{\scriptscriptstyle 0}}

\def\muo{\mu_{\scriptscriptstyle 0}}

\def\ko{k_{\scriptscriptstyle 0}}
\def\co{c_{\scriptscriptstyle 0}}

\def\ux{\hat{\#u}_x}
\def\uy{\hat{\#u}_y}

\def\QL{\#Q_{\rm L}}
\def\QR{\#Q_{\rm R}}
\def\kL{k_{\rm L}}
\def\kR{k_{\rm R}}
\def\kLBr{k_{\rm L}^{\rm Br}}
\def\kRBr{k_{\rm R}^{\rm Br}}

\def\epsa{\eps^{\rm a}}
\def\epsb{\eps^{\rm b}}
\def\epsBr{\eps^{\rm Br}}
\def\mub{\mu^{\rm b}}
\def\muBr{\mu^{\rm Br}}
\def\xib{\xi^{\rm b}}
\def\xiBr{\xi^{\rm Br}}

\def\fa{f_{\rm a}}
\def\fb{f_{\rm b}}

\begin{document}

\begin{center}

\LARGE{ {\bf Simultaneous amplification and attenuation in isotropic chiral materials
}}
\end{center}

\begin{center}
\vspace{10mm} \large

 Tom G. Mackay\footnote{E--mail: T.Mackay@ed.ac.uk.}\\
{\em School of Mathematics and
   Maxwell Institute for Mathematical Sciences\\
University of Edinburgh, Edinburgh EH9 3FD, UK}\\
and\\
 {\em NanoMM~---~Nanoengineered Metamaterials Group\\ Department of Engineering Science and Mechanics\\
Pennsylvania State University, University Park, PA 16802--6812,
USA}\\
 \vspace{3mm}
 Akhlesh  Lakhtakia\footnote{E--mail: akhlesh@psu.edu}\\
 {\em NanoMM~---~Nanoengineered Metamaterials Group\\ Department of Engineering Science and Mechanics\\
Pennsylvania State University, University Park, PA 16802--6812, USA}

\normalsize

\end{center}

\begin{center}
\vspace{15mm} {\bf Abstract}

\end{center}

The electromagnetic field phasors in
an isotropic chiral material (ICM) are superpositions of two Beltrami fields of different
handedness. Application of the Bruggeman homogenization formalism to two-component composite materials delivers ICMs
 wherein Beltrami fields of one handedness attenuate whereas Beltrami waves of the other handedness amplify.
One component material is a dissipative ICM, the other  an active dielectric material.

\vspace{5mm}


\vspace{14mm}

The issues of gain and loss  are currently prominent ones  in
electromagnetics, as
 active component materials are being introduced in electromagnetic metamaterials
 in order to overcome losses \c{Wuestner,Dong_APL,Strangi}. Depending on the
imaginary part of its    permittivity scalar  $\eps(\omega)$ at angular frequency $\omega$,
an isotropic dielectric material is  either (i) dissipative if ${\rm Im} [\eps(\omega)] > 0$  or (ii) active if  ${\rm Im}  [\eps(\omega)]  < 0$ or (iii) neither
 if ${\rm Im} [\eps(\omega)] =0$, provided that the electromagnetic fields are assumed
 to depend as  $\exp \le - i \omega t \ri$  on time $t$
\c{Chen}.
Whether an anisotropic dielectric material is dissipative or active
 is determined by the imaginary  part of its permittivity dyadic $\=\eps(\omega)$ \c{Tan}. Furthermore, anisotropic dielectric materials for which the imaginary part of $\=\eps(\omega)$ is indefinite can exhibit gain for certain propagation directions and loss for other propagation directions \c{ML_PRA}.

This short article concerns the issue of simultaneous attenuation and amplification during planewave propagation along any specific direction in an isotropic chiral material (ICM). Such a material is
 characterized by  the frequency-domain Tellegen constitutive relations \c{Beltrami}\footnote{Here, and henceforth, the dependencies of the constitutive parameters and field phasors on $\omega$ are not  explicitly displayed.}
\begin{equation} \l{cr}
\left.
\begin{array}{l}
 \#D (\#r) = \eps\#E (\#r) + i \xi \#H (\#r) \,\\ [5pt]
 \#B (\#r)= - i \xi \#E (\#r) + \mu \#H (\#r) \,
\end{array}
\right\}\,.
\end{equation}
The scalar constitutive parameters $\eps$, $\xi$, and $\mu$ are complex-valued in general.
The Bohren decomposition
\begin{equation}
\left.
\begin{array}{l}
\#E(\#r)= \QL(\#r)-i\eta\QR(\#r)\,\\[5pt]
\#H(\#r)=(i\eta)^{-1}\QL(\#r)+\QR(\#r)
\end{array}
\right\}
\end{equation}
is employed to represent $\#E$ and $\#H$ as superpositions of a left-handed Beltrami
field $\QL$ and a right-handed Beltrami field $\QR$, with $\eta = \mu^{1/2}\eps^{-1/2}$ \cite{Beltrami}. In source-free regions, the two Beltrami fields obey the relations
\begin{equation}
\left.
\begin{array}{l}
\nabla\times\QL(\#r)= \kL\QL(\#r)\\[5pt]
\nabla\times\QR(\#r)= -\kR\QR(\#r)
\end{array}
\right\}\,,
\end{equation}
where the wavenumbers
\begin{equation} \l{wavenumbers}
\left.
\begin{array}{l}
\kL= \omega\left(\mu^{1/2}\eps^{1/2}+\xi\right)\\[5pt]
\kR= \omega\left(\mu^{1/2}\eps^{1/2}-\xi\right)
\end{array}
\right\}\,.
\end{equation}

Let us consider planewave propagation along the $+z$ direction. Then $\QL=(\ux+i\uy)\exp\left(i\kL{z}\right)$ is a left-circularly polarized (LCP) plane wave and $\QR=(\ux-i\uy)\exp\left(i\kR{z}\right)$ is a right-circularly polarized (RCP) plane wave. Could the LCP plane wave lose energy and the RCP plane wave gain energy, or \textit{vice versa}, as $z$ increases?  More generally, could Beltrami waves of one handedness attenuate while Beltrami waves of the other handedness amplify, even if these Beltrami waves are not plane waves, but, say, spherical or cylindrical waves?
In other words,
could ${\rm Im}\left(\kL\right)$ and ${\rm Im}\left(\kR\right)$ be of opposite signs, but
${\rm Re}\left(\kL\right)$ and ${\rm Re}\left(\kR\right)$ have the same signs?  If yes,
then ICM research is promising for circular polarizers of a new type.

A perusal of literature on ICMs did not turn up any example for which  ${\rm Im}\left(\kL\right){\rm Im}\left(\kR\right) < 0$ but ${\rm Re}\left(\kL\right) {\rm Re}\left(\kR\right) > 0$. Hence, we decided to investigate a particulate composite material   comprising   an active dielectric material and a dissipative  ICM. If the component materials can be regarded as being randomly distributed as electrically small particles, then the  composite material could be homogenized into an ICM itself \cite{MAEH}.

Let the component material labeled `a' be an active isotropic dielectric material specified by
 the permittivity $\epsa$ such that ${\rm Re}\le \epsa\ri>0$ and ${\rm Im}\le \epsa\ri<0$.
Let the component material `b' be a dissipative ICM,
characterized by constitutive relations of the form given in  Eqs.~\r{cr}, but with the superscript `b' attached to the
constitutive parameters $\eps$, $\xi$, and $\mu$ therein.
We used the well-established Bruggeman formalism \cite{Kampia} to estimate the
 constitutive parameters $\epsBr$, $\xiBr$, and $\muBr$ of the homogenized
 composite material (HCM),
 per  Eqs.~\r{cr} but with the superscript `Br' attached to the
constitutive parameters $\eps$, $\xi$, and $\mu$ therein.
 Let
  $\fa\in[0,1]$ denote  the volume fraction of component material `a', the volume fraction of
  component material `b' being $\fb=1-\fa$.

\begin{figure}[!htb]
\begin{center}
\begin{tabular}{c}
\includegraphics[width=14.0cm]{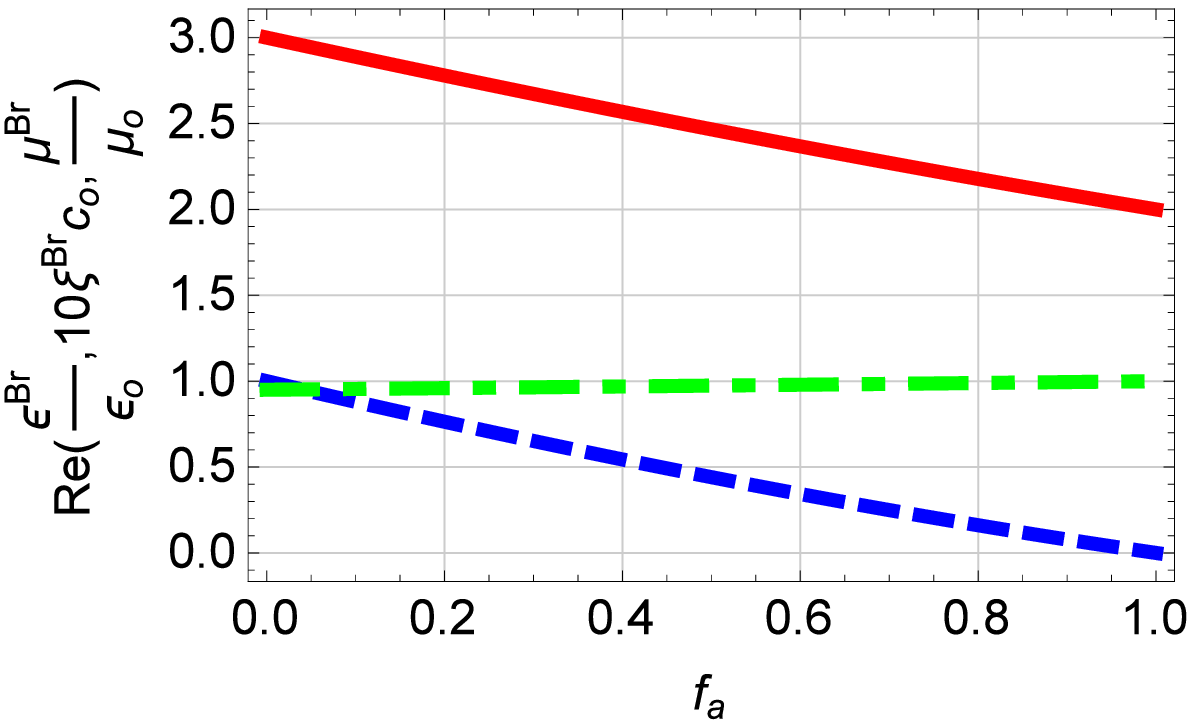}\\
\includegraphics[width=14.0cm]{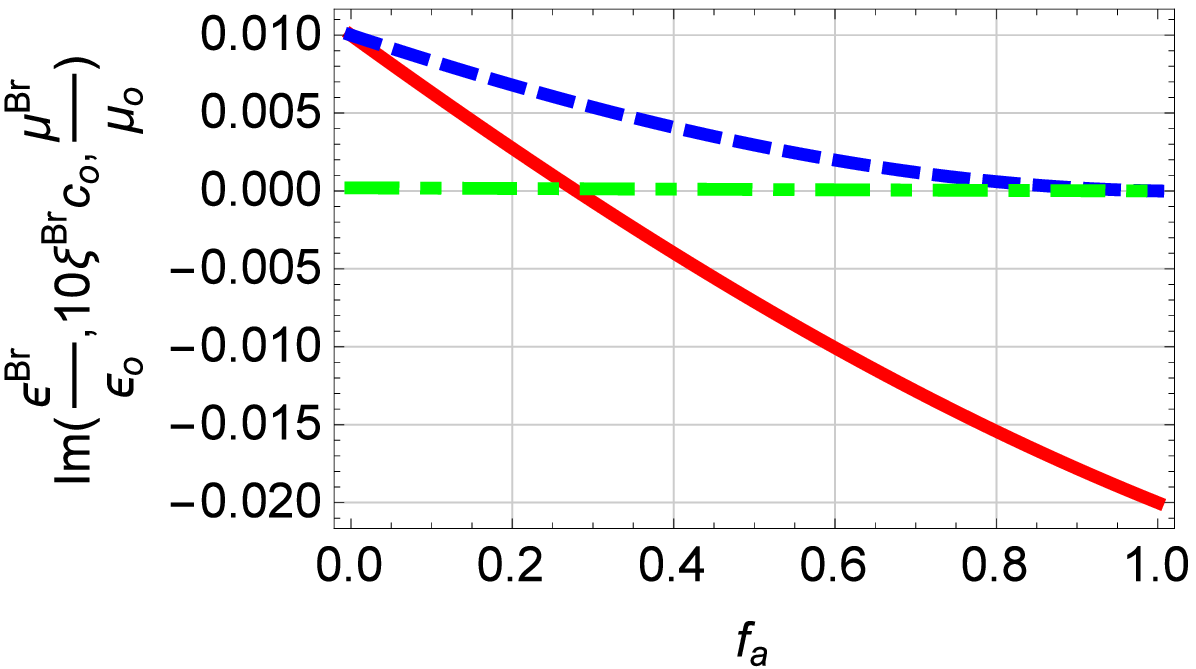}
\end{tabular}
\end{center}
 \caption{The real and imaginary parts of the
 constitutive parameters
   $\epsBr/\epso$  (red solid curves), $\xiBr \co$ (blue dashed curves), and $\muBr / \muo$ (green broken-dashed curves) of the HCM, as estimated using the Bruggeman formalism, plotted against volume fraction $\fa$. The real and imaginary parts of $\xiBr \co$ are scaled by a factor of 10. See the text for values of $\epsa$, $\epsb$, $\xib$, and $\mub$.
  } \label{fig1}
\end{figure}

Figure~\ref{fig1}  shows the   real and imaginary parts of $\epsBr$, $\xiBr$, and $\muBr$ as functions of $\fa$, when
 $\epsa = \le 2.0 - 0.02 i \ri \epso$,
  $\epsb = \le 3 +  0.01 i \ri \epso$,  $\xib = \le 0.1 + 0.001 i \ri/\co  $,
   and  $\mub = \le 0.95 + 0.0002 i \ri \muo$, with $\epso$ and $\muo$ being the permittivity and  permeability of free space, respectively, and $\co=  1/\sqrt{\epso \muo}$.
The component material `b' is guaranteed to be dissipative since $  \les  \mbox{ Im} \le \xib \ri \ris^2 <
 \mbox{ Im} \le \epsb \ri \mbox{ Im} \le \mub \ri $
\c{Lindell}.
The chosen values of $\epsa$, $\epsb$, $\xib$, and $\mub$ are physically plausible
\cite{ML_PRA,Sun,Gomez}.

The real and imaginary parts of $\epsBr$, $\xiBr$, and $\muBr$ vary almost linearly in Fig.~\ref{fig1} as $\fa$ increases from 0 to 1, with their endpoints complying to the limits
\begin{equation}
\epsBr \to \left\{ \begin{array}{l} \epsa \\  \epsb \end{array} \right. , \quad \xiBr \to \left\{ \begin{array}{l} 0 \\  \xib \end{array} \right. , \quad
\muBr \to \left\{ \begin{array}{l} 1 \\  \mub \end{array} \right. \quad \mbox{as} \quad
\fa \to \left\{ \begin{array}{l} 1 \\  0 \end{array} \right. .
\end{equation}
In particular,  the sign of $\mbox{Im} \le \epsBr \ri$ changes at  $\fa \approx 0.27$.

\begin{figure}[!htb]
\begin{center}
\begin{tabular}{c}
\includegraphics[width=14.0cm]{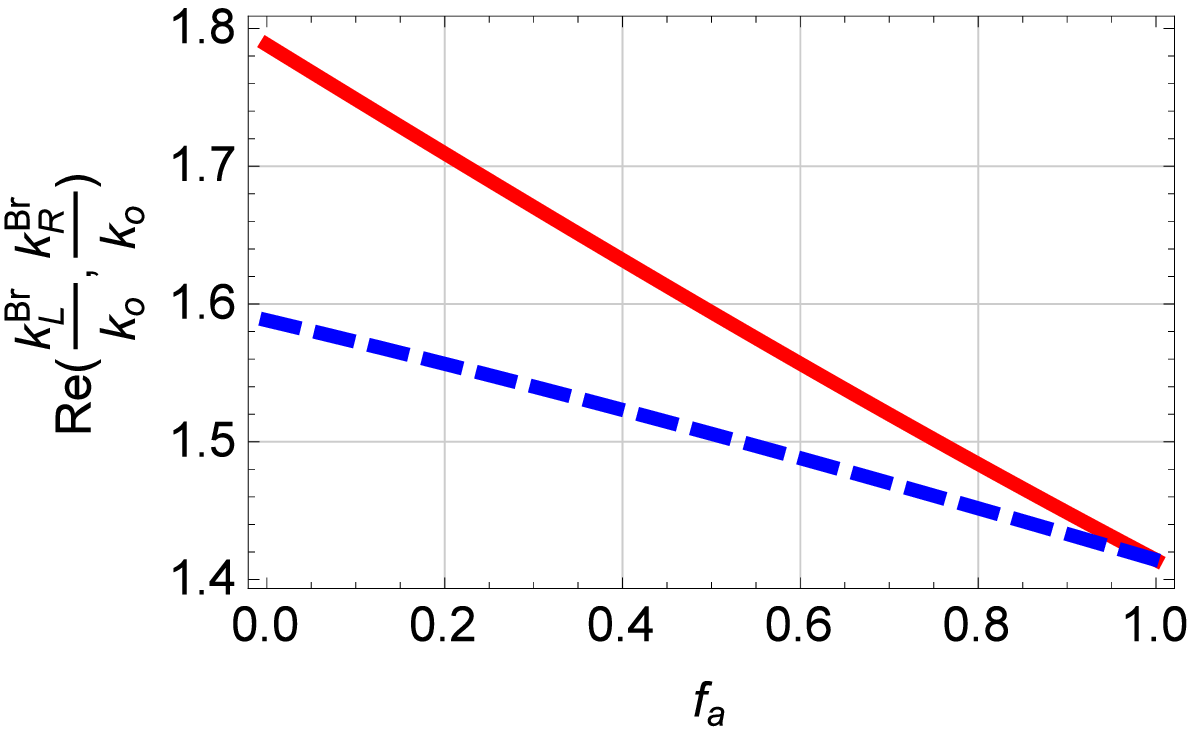}\\
\includegraphics[width=14.0cm]{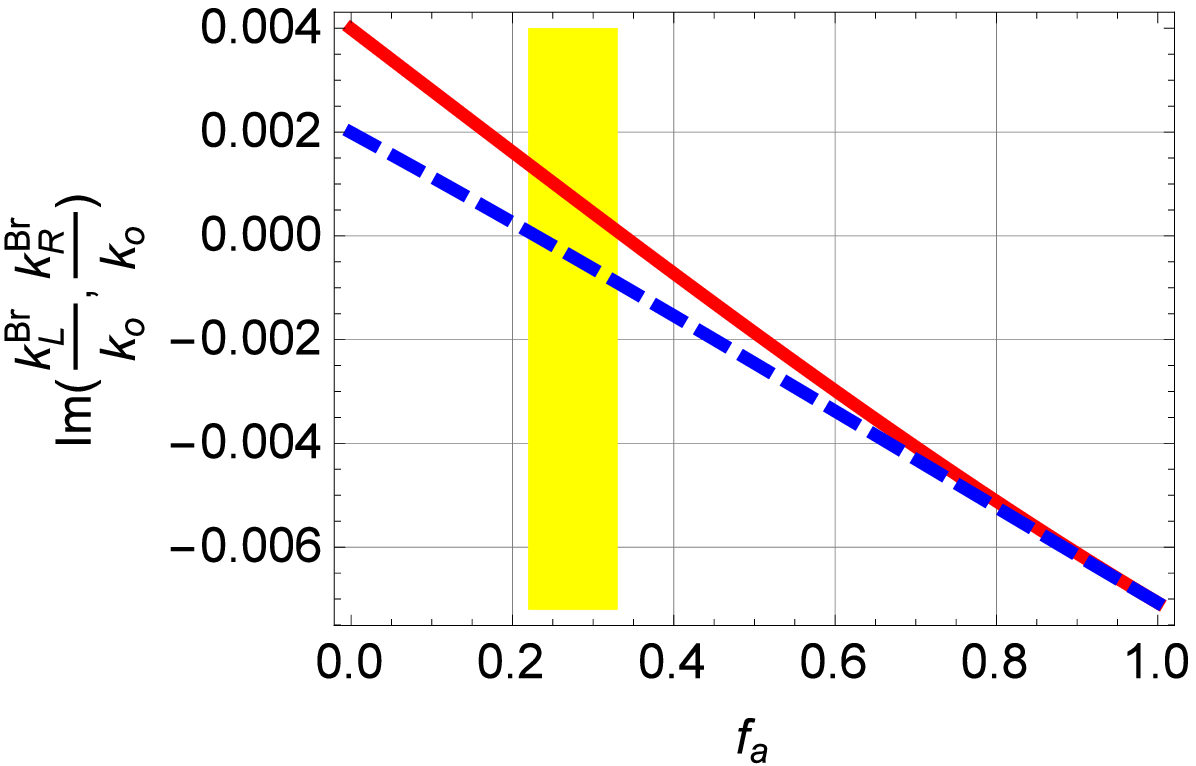}
\end{tabular}
\end{center}
 \caption{The real and imaginary parts of the  relative wavenumbers $\kLBr$  (red solid curves)
and $\kRBr$  (blue dashed curves) in the HCM, normalized with respect to the free-space wavenumber
$\ko=\omega/\co$, plotted against volume fraction $\fa$.     The component materials are the same as for Fig.~\ref{fig1}. The volume fraction range where $
\mbox{Im} \left(\kLBr\right) \mbox{Im} \left(\kRBr\right) < 0$ is shaded in yellow.
 } \label{fig2}
\end{figure}

The real and imaginary parts of the   wavenumbers for the HCM---namely $\kLBr$
and $\kRBr$ per Eqs.~\r{wavenumbers} with the superscript `Br' attached to $\kL$, $\kR$, $\eps$, $\xi$, and $\mu$ therein---are plotted against volume fraction $\fa$ in Fig.~\ref{fig2}. Whereas
${\rm Re}\left(\kLBr\right)>0$ and ${\rm Re}\left(\kRBr\right)>0$ for any $\fa\in[0,1]$,
three mutually disjoint $\fa$-regimes can be identified for the signs of
${\rm Im}\left(\kLBr\right)$ and ${\rm Im}\left(\kRBr\right)$ as follows:
\begin{itemize}
\item[(i)]  {$\fa\in[0,0.22)$}, when ${\rm Im}\left(\kLBr\right)>0$ and ${\rm Im}\left(\kRBr\right)>0$;
\item[(ii)]  {$\fa\in(0.22,0.33)$}, when ${\rm Im}\left(\kLBr\right)>0$ and ${\rm Im}\left(\kRBr\right)<0$; and
\item[(iii)] $\fa\in(0.33,1]$, when ${\rm Im}\left(\kLBr\right)<0$ and ${\rm Im}\left(\kRBr\right)<0$.
\end{itemize}
Clearly then, a continuous range of values of the volume fraction $\fa$, specifically $\fa\in(0.22,0.33)$ for the example
presented in Fig.~\ref{fig2}, can exist wherein  $\QL$  attenuates whereas  $\QR$  amplifies.

\begin{figure}[!htb]
\begin{center}
\begin{tabular}{c}
\includegraphics[width=14.0cm]{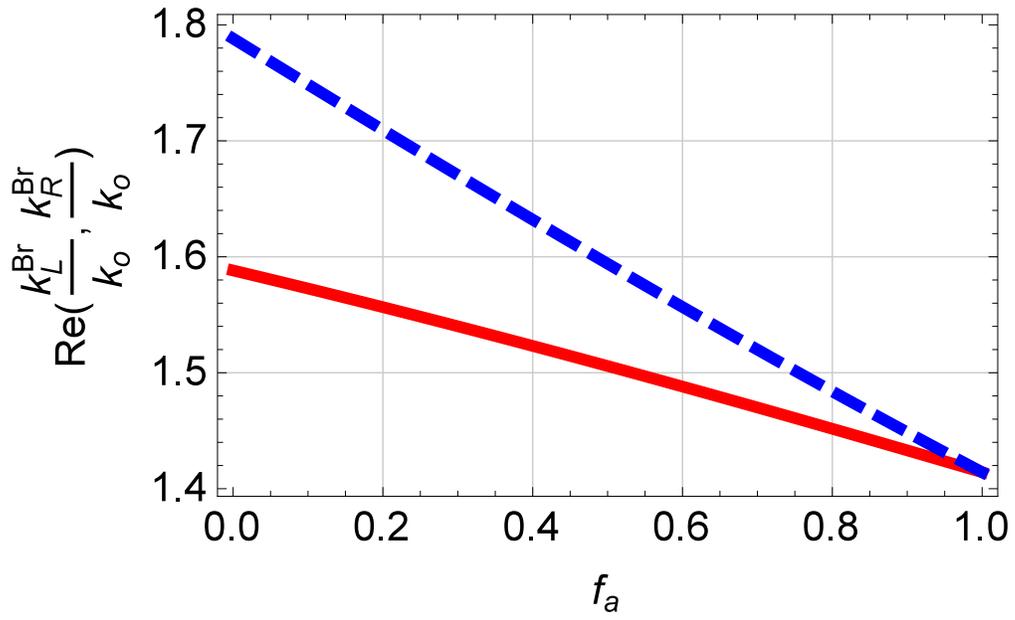}\\
\includegraphics[width=14.0cm]{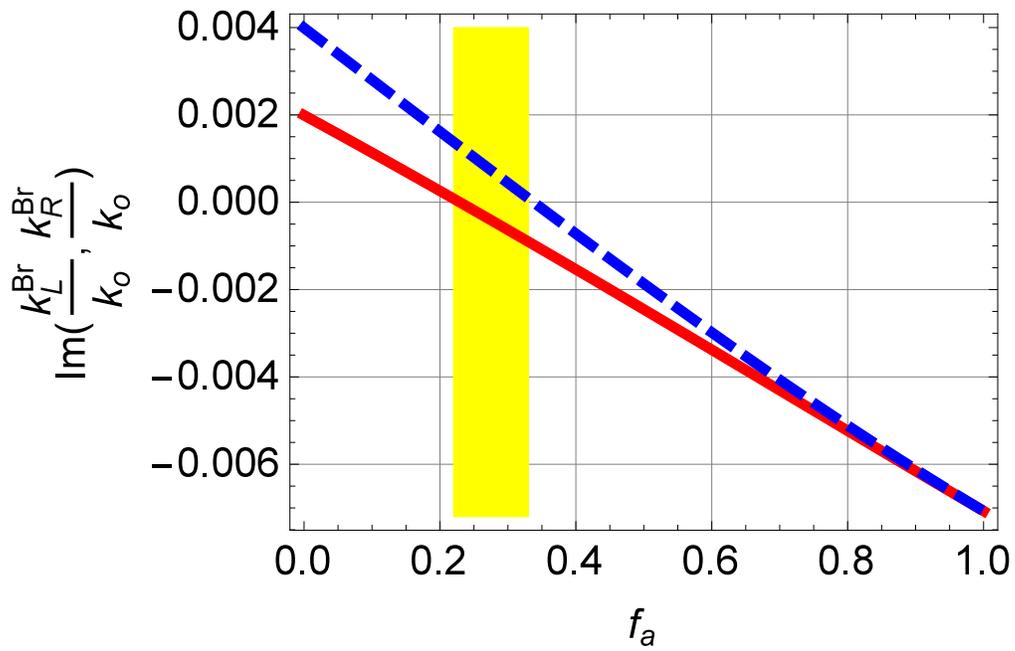}
\end{tabular}
\end{center}
 \caption{As Fig.~\ref{fig2} but with $\xib$ replaced by $-\xib$.
 } \label{fig3}
\end{figure}

If $\xi$ is replaced by $-\xi$ in Eqs.~\r{wavenumbers}, then $k_L$ and $k_R$ in Eqs.~\r{wavenumbers} are interchanged.
Also, if the sign of $\mbox{Re} \le \xi \ri $ is reversed in Eqs.~\r{wavenumbers}, then $\mbox{Re} \le k_L \ri$ and $\mbox{Re} \le k_R \ri$  are interchanged, but $\mbox{Im} \le k_L \ri$ and $\mbox{Im} \le k_R \ri$ remain unchanged. Therefore, if  $\xib$ were to be replaced by $-\xib$ for the homogenization scenario represented in Figs.~\ref{fig1} and \ref{fig2}, then the handedness of the Beltrami field that is amplified/attenuated for the regime wherein
 ${\rm Im}\left(\kLBr\right){\rm Im}\left(\kRBr\right) < 0$ will be reversed. To illustrate this point, in Fig.~\ref{fig3}
 the wavenumbers  $\kLBr$
and $\kRBr$  are plotted against  $\fa$ for the same homogenization scenario as represented in Figs.~\ref{fig1} and \ref{fig2} but with  $\xib = - \le 0.1 + 0.001 i \ri/\co  $. For $\fa\in(0.22,0.33)$, we infer from Fig.~\ref{fig3} that  $\QL$  is amplified and  $\QR$   is attenuated, whereas from Fig.~\ref{fig2} it is  $\QR$  that is amplified and  $\QL$  that is attenuated.

In the foregoing analysis, a physically plausible means of achieving an ICM for which
left-handed Beltrami fields are amplified whereas right-handed Beltrami fields are attenuated (or \textit{vice versa}) is conceptualized. This result opens the door
 for circular polarizers of a new type.

\vspace{5mm}

\noindent {\bf Acknowledgment:} TGM acknowledges the support of EPSRC grant EP/M018075/1. 
AL thanks the Charles Godfrey Binder Endowment at Penn State for ongoing  support of his
research activities.

\newpage

\newpage

\newpage

\end{document}